\documentclass{eptcs}

\usepackage{amssymb}
\usepackage{t1enc}
\usepackage[latin1]{inputenc}
\usepackage{amsmath}
\usepackage{mathrsfs}
\usepackage{stmaryrd}
\usepackage{graphicx}
\usepackage{boxedminipage}
\usepackage{algpseudocode}
\usepackage{listings}
\usepackage[inline,nomargin,draft]{fixme}
\usepackage{xspace}
\usepackage{url}
\usepackage{color} 
\usepackage{enumerate}
\usepackage{mdwlist}

\begin{document}

\newcommand\mono[1]{\texttt{#1}}

\providecommand{\event}{LAFM 2013}

\title{On Verifying Resource Contracts using Code Contracts}
\def\titlerunning{On Verifying Resource Contracts using Code Contracts}

\author{
Rodrigo Casta\~no,  Diego Garbervetsky, \\ Jonathan Tapicer, Edgardo Zoppi
\institute{Departamento de Computaci\'on. FCEyN. UBA\\Buenos Aires, Argentina}
\email{\{rcastano, diegog, jtapicer, ezoppi\}@dc.uba.ar}
\and
Juan Pablo Galeotti 
\institute{Saarland University \\ Saarbr\"ucken, Germany}
\email{galeotti@cs.uni-saarland.de}
}

\def\authorrunning{R. Casta\~no, J. P. Galeotti, D. Garbervetsky, J. Tapicer, E. Zoppi}
\maketitle


\begin{abstract}
In this paper we present an approach to check resource consumption contracts using an off-the-shelf static analyzer.
We propose a set of annotations to support resource usage specifications, in particular,  dynamic memory consumption constraints.
Since dynamic memory may be recycled by a memory manager, the consumption of this resource is not monotone.
The specification language can express both memory consumption and lifetime properties in a modular fashion.
We develop a proof-of-concept implementation by extending \textsc{Code Contracts}' specification language. 
To verify the correctness of these annotations we rely on the \textsc{Code Contracts} static verifier and a points-to analysis.
We also briefly discuss possible extensions of our approach to deal with non-linear expressions.
\end{abstract}


\section{Problem statement}
Design by contract~\cite{Meyer:OOP} is a programming discipline that prescribes that software designers should define formal, precise and verifiable interface specifications for software components, extending the ordinary definition of abstract data types with preconditions, postconditions and invariants.  

While there has been some success in the adoption of contracts for enforcing functional requirements and design decisions~\cite{leavens00jml,Meyer:Eiffel}, there have not been many signs of their usage to express non-functional requirements such as performance or resource utilization requirements.  
Possible causes are the inherent difficulty of writing quantitative requirements, the lack of a convenient language to express them and tool support to verify them. 
However, in many settings it is crucial to enforce the fulfillment of this kind of requirements. Certifying memory consumption is vital to ensure safety in embedded systems; understanding the number of messages sent through a network is useful to detect performance bottlenecks or reduce communication costs, etc. 
It is well known that inferring, and even checking, quantitative bounds (e.g., resource usage) is  difficult~\cite{garber08ismm}. Nevertheless, there has been noticeable progress in techniques that compute symbolic resource usage~\cite{garber08ismm, AlbertGG09} and complexity~\cite{gulwani2010reachability} upper-bounds.

In this paper we focus in enforcing dynamic memory consumption contracts. 
This is a particularly challenging problem because memory footprint does not monotonically increase during program execution (i.e., due to memory recycling). For programming languages with automatic memory reclaiming mechanisms (such as .NET based languages), this problem gets even more complex since memory consumption depends on the behavior of both the application and the garbage collector (GC). 
We believe other quantitative requirements may be computed by using a similar approach. 

\textsc{Code Contracts}~\cite{fahndrich2010embedded} is a tool that brings the advantages of design-by-contract programming to all .NET based programming languages enabling the use of contracts without requiring a specific compiler. 

We present an extension of the \textsc{Code Contracts} annotation language designed to specify the amount of memory consumed by a method. These specifications have two possible interpretations:  on one hand, they state that a method consumes less memory than a particular bound but, once verified, they can be interpreted as well as a precondition stating the system requires at least the amount of memory specified in order to run safely.

The proposed extension also provides means for specifying object lifetimes needed to model object allocation and reclaiming. Roughly speaking, we distinguish between \emph{temporary} and \emph{escaping} objects. 
Escaping objects refer to objects that are created  in a method but may be used by its callers and, thus,  should live longer.
Temporary objects can be safely collected once the method finishes its execution. 
We provide constructs to enable client methods to reclaim some of these objects. For verification purposes, it is important to provide an upper bound to these objects. By doing so, it is possible to use these bounds to verify each method's contract separately.

In order to verify the annotations we instrument the original program, adding special counters and translating the annotations to conventional statements in terms of these counters.
In general, we are assuming there is a static verification engine for the language in which the original program was coded.
In our implementation, we rely on \textsc{Clousot}~\cite{clousot}, the static analyzer used by \textsc{Code Contracts}, which is based in abstract interpretation~\cite{Cousot77}.
One of the main advantages of \textsc{Clousot} is that it is able to verify programs featuring loops without the need of loop invariants.
Its main drawback is that it cannot handle non-linear expressions, which may appear in complex programs.  In the paper we discuss some possible approaches to overcome this problem.

All this work has been implemented as a \textsc{Visual Studio} plugin enabling static verification and run-time checks of memory consumption contracts.

The paper is organized as follows: in~\S\ref{sec:annotations} we introduce a set of annotations to describe memory consumption  contracts,  objects' lifetime information and iteration spaces for loops. In~\S\ref{sec:verification} we show how to transform those annotations into code and contracts supported by \textsc{Clousot}, and how we check object lifetime annotations.
In~\S\ref{sec:barvinok} we discuss one possible  extension of the checker in order to support polynomial constraints using the \textsc{Barvinok} calculator.
Then, in~\S\ref{sec:implementation} we present some implementation details. We conclude in~\S\ref{sec:relatedwork} and~\S\ref{sec:conclusions} discussing some related and future work.

\section{Memory usage annotations} \label{sec:annotations}
In this section  we present our annotation language.
Its design was driven by the following considerations: 
\begin{enumerate}[(i)]
	\item The annotations need to provide means to specify that objects are allocated but also potentially reclaimed by the GC in a simple and modular fashion (lifetime information). 
	\item The annotations should be rich enough to allow client methods to check its own annotations using the callees' resource specifications without losing much precision. 
	\item Both quantitative and lifetime constraints have to be in terms of method parameters and instance variables.
	\item The mechanism to specify consumption information should  maintain certain basic encapsulation properties such as information hiding.
\end{enumerate}

To represent memory recycling due to GC we based our annotation language on a very simple memory model\footnote{This model is inspired in the scoped-memory management proposed for Real-Time Java~\cite{bollella00realtime}, but in this case we just used it as an over-approximation of GC behavior.}: objects can be annotated as \emph{escaping}, meaning that they may be used by a client method and, therefore, should live longer than the method itself, or not annotated, meaning that they are used only for auxiliary computation and are no longer needed at the end of the method execution. Furthermore, we only allow annotations to express properties about objects created in the scope of the method in which they are contained.
The annotations are based on escape analysis terminology, escaping objects escape the method's scope, whereas all other objects are captured by the method.

\mono{MemReq} is used to specify an upper bound of the maximum number of  \emph{live} objects that were allocated  by the method.
By live objects we mean the objects  that could not be collected by the garbage collector during the execution of the method.
In what follows, unless explicitly stated, by memory requirements we refer to \mono{MemReq}.
\mono{Esc}  specifies an upper bound to the total number of allocated objects escaping the method. 
These annotations must be placed at the beginning of a method. They expect a class name and an integer expression which declares the number of objects of that class consumed by the method.
Notice that these annotations should be interpreted within the method as an ensures clause stating that the method consumes at most the declared number of objects, but from the client point of view its role is a requires clause demanding that the system needs at least that space for the specified quantity of objects in order to safely run. We will also use the first interpretation to introduce a compositional verification algorithm. 

Figure~\ref{exMemReq} shows a brief example, written in C\#, of the \mono{MemReq} annotation.
The \mono{Person} constructor creates a temporary \mono{Logger} which is discarded and not reachable from outside the scope of the method.
This object is captured by the method, therefore it is not annotated as escaping. 

\begin{figure}[ht]
\begin{scriptsize}
\begin{lstlisting}
class Person {
	public string FirstName { get; private set; }
	public string LastName { get; private set; }

	public Person(string firstName, string lastName) {
		Contract.Memory.MemReq<Logger>(1);

		this.FirstName = firstName;
		this.LastName = lastName;

		Logger logger = new Logger();
		logger.logMessage("Person created.");
	}
}
\end{lstlisting}
\end{scriptsize}
\vspace{-.5em}
\caption{Annotated method featuring only captured objects.}
\label{exMemReq}
\end{figure}

In addition to the quantitative expression, \mono{Esc} expects an identifier for tagging this set of escaping objects. The tag is used to specify that those objects belong to a group having similar characteristics in terms of lifetime (e.g., they are part of the same data structure). For instance, the identifier \mono{Contract.Memory.Return} indicates that this set of objects is returned and  \mono{Contract.Memory.This} indicates that this set of objects may be reachable by the receiver. A developer can define an arbitrary set of identifiers according to her needs of distinguishing sets of escaping objects.

To verify the aforementioned contracts we need to inform the lifetime of every object allocated by the method.  To do so, we introduce a new annotation: \mono{DestEsc}, which should be located before every \mono{new} statement.
\mono{DestEsc(t)} declares an object as escaping (living longer that the method itself) and associates the object with one of the tags already mentioned in the contract.
If the annotation were missing, the object will be considered temporary, and it should not be accessible from outside the scope of the method.

The code of the \mono{Family} constructor, shown in Figure~\ref{exEsc}, includes an escaping object.
As can be seen, the contract of line \ref{exEsc:esc} states that an array of type \mono{Person} is escaping the scope of the method. Instead of representing the array as one object, we quantify it by its size. 
Additionally, the \mono{new Person[size];} statement of line \ref{exEsc:newPerson} is tagged with \mono{DestEsc(Contract.Memory.This)}, indicating that the escaping object will be reachable from outside the scope of the method through the receiver of the method call (in this case, since the method is a constructor, the receiver is the newly created \mono{Family} instance).
The method \mono{AddMember} creates an object of type \mono{Person} and includes it in the array of \mono{Family} instances.
Since it calls the constructor of \mono{Person}, it requires space for one object of type \mono{Logger}  and one of type \mono{Person}.
This last object escapes the scope of the method because it is referenced by the instance field \mono{\_Members}. Thus, we include the annotation \mono{DestEsc(Contract.Memory.This)} of line \ref{exEsc:destEsc}.  

\begin{figure}[ht]
\begin{scriptsize}
\begin{lstlisting}[escapeinside={@}{@}]
class Family {
	private Person[] _Members;
	public string LastName { get; private set; }
	public int Size { get; private set; }

	public Family(string lastName, int size) {
		Contract.Memory.MemReq<Person[]>(size);
		@\label{exEsc:esc}@Contract.Memory.Esc<Person[]>(Contract.Memory.This, size);

		Contract.Memory.DestEsc(Contract.Memory.This);
		@\label{exEsc:newPerson}@_Members = new Person[size];

		this.LastName = lastName;
		this.Size = 0;
	}

	public void AddMember(string firstName) {
		Contract.Requires(this.Size < _Members.Length);
		Contract.Memory.MemReq<Person>(1);
		Contract.Memory.MemReq<Logger>(1)
		Contract.Memory.Esc<Person>(Contract.Memory.This, 1);

		@\label{exEsc:destEsc}@Contract.Memory.DestEsc(Contract.Memory.This);
		Person person = new Person(firstName, this.LastName);
		_Members[this.Size++] = person;
	}
}
\end{lstlisting}
\end{scriptsize}
\vspace{-.5em}
\caption{Annotated method featuring escaping objects.}
\label{exEsc}
\end{figure}

The example in Figure~\ref{exMCall} brings together most of the annotations shown so far.
The statement in line \ref{exMCall:newFamily} shows an example of objects escaping from a method call. In this case, the escaping objects are associated with the returned object. In fact, the \mono{Family} returned will be precisely the one created in that statement. 
In the case of method invocations we need to figure out the destination of escaping objects originated in callees. 
The annotation \mono{AddEsc(dst, src)} states that objects escaping the callee, tagged with \mono{src}, become also escaping objects of the caller, but identified with the tag \mono{dst}.
All objects escaping a method invocation and missing the corresponding annotation will be considered temporary for the caller and should not be accessible from outside its scope.

It is important to mention the reason for the difference between the total memory requirements and the number of escaping objects.
Consider the call to  \mono{AddMember}.  As specified in Figure~\ref{exMemReq}, a temporary \mono{Logger} will be created by the \mono{Person} constructor. Assuming the contract to be correct, the memory requirements of the callee, in this case, have to be added to the total memory requirements of the method \mono{CreateFamily}, but since this is a temporary object, the memory required for this object can be recycled after the call to \mono{AddMember}. 
In contrast, escaping objects are accumulative meaning that the caller needs space for each object escaping from \mono{AddMember} at each iteration.


\begin{figure}[ht]
\begin{scriptsize}
\begin{lstlisting}[escapeinside={@}{@}]
Family CreateFamily(string lastName, string[] firstNames) {
	Contract.Memory.MemReq<Family>(1);
	Contract.Memory.MemReq<Person[]>(firstNames.Length);
	Contract.Memory.MemReq<Person>(firstNames.Length);
	Contract.Memory.MemReq<Logger>(1);
	Contract.Memory.Esc<Family>(Contract.Memory.Return, 1);
	Contract.Memory.Esc<Person[]>(Contract.Memory.Return, firstNames.Length);
	Contract.Memory.Esc<Person>(Contract.Memory.Return, firstNames.Length);

	Contract.Memory.DestEsc(Contract.Memory.Return);
	Contract.Memory.AddEsc(Contract.Memory.Return, Contract.Memory.This);
	@\label{exMCall:newFamily}@Family family = new Family(lastName, firstNames.Length);

	for (int i = 0; i < firstNames.Length; ++i) {
		Contract.Memory.AddEsc(Contract.Memory.Return, Contract.Memory.This);        
		family.AddMember(firstNames[i]);
	}

	return family;
}
\end{lstlisting}
\end{scriptsize}
\vspace{-.5em}
\caption{Annotated method featuring most annotations presented.}
\label{exMCall}
\end{figure}

So far, we have been using tags to declare sets of escaping objects.
This mechanism encompasses information hiding and is sufficient to specify and enforce the quantitative aspects of method consumption.
However, to check the validity of annotations concerning objects' lifetime, namely \mono{DestEsc}, we need to provide the checker with the means to link tags to actual objects.
To do that, we introduce the annotation \mono{BindEsc(t, expr)} which connects a tag \mono{t} with a set of objects referred by the path-expression \mono{expr}.
For instance, in line \ref{exBind:bindEsc} of Figure~\ref{exBind}, \mono{BindEsc(Param, brother)} specifies that the tag \mono{Param} represents all objects reachable from the returning parameter  \mono{brother}. 

\begin{figure}[ht]
\begin{scriptsize}
\begin{lstlisting}[escapeinside={@}{@}]
Person CreateBrothers(string lastName, string name1, string name2, out Person brother) {
	Contract.Memory.MemReq<Person>(2);
	Contract.Memory.MemReq<Logger>(1);
	Contract.Memory.Esc<Person>(Contract.Memory.Return, 1);
	Contract.Memory.Esc<Person>(Param, 1);
 	@\label{exBind:bindEsc}@Contract.Memory.BindEsc(Param, brother);
    
	Contract.Memory.DestEsc(Contract.Memory.Return);
	Person person = new Person(name1, lastName);

	Contract.Memory.DestEsc(Param);
	brother = new Person(name2, lastName);
	return person;
}
\end{lstlisting}
\end{scriptsize}
\vspace{-.5em}
\caption{Using \mono{BindEsc} to relate tags with path expressions.}
\label{exBind}
\end{figure}

It is worth noticing that \mono{AddEsc}, \mono{DestEsc} and \mono{BindEsc} are internal method annotations, not visible outside the method boundary. In contrast, \mono{MemReq}, \mono{Esc} and their tags can be used by clients.

\section{Verifying memory consumption} \label{sec:verification}
To automatically check the annotations introduced in the previous section we transform the annotated program into a functionally equivalent instrumented program in such a way  that a successful  verification of the transformed program implies the correctness of the original resource usage annotations.
For the case of .NET programs, part of the transformation involves we rely on \textsc{Code Contracts}' basic annotations.
The instrumentation is performed at the IL level and is never read or manipulated by developers.

\subsection{Introducing counters and ensure clauses}

In order to use standard notation, we transform each memory consumption restriction into assertions in terms of integer counters.
With that purpose in mind, we will keep track of one counter for each memory lifetime tag and an additional one for total memory requirements of objects of a particular type.

For every method \mono{m} featuring memory consumption we apply the following procedure:

Let $T$ be the set of memory lifetime tags appearing in the contract of \mono{m} and $C$ the set of classes. For each tag $\mono{t} \in T$ and $\mono{C} \in C$ we introduce the counter \mono{m\_Esc\_t\_C}. We will also create the counter \mono{m\_MemReq\_C}.
The first counter tracks the number of objects of type \mono{C} escaping from \mono{m}  that are associated with tag \mono{t}.
In contrast, the latter is incremented with every object creation in \mono{m}, since it represents the total memory requirements of \mono{m} for objects of type $\mono{C}$.
To keep the counters updated, \mono{m\_MemReq\_C} is incremented before each \mono{new C()} statement in \mono{m}. If the statement were annotated with \mono{DestEsc}, we also introduce a statement to increment \mono{m\_Esc\_t\_C}. 

To illustrate the idea consider the example in Figure~\ref{ex-baseinst} which allocates two objects of type \mono{A} and performs two method calls. 
Figure~\ref{ex-inst} shows the instrumented version where the counters 
\mono{m\_MemReq\_A}, \mono{m\_Esc\_Return\_A} and \mono{m\_Esc\_Param\_A} track respectively the amount of objects of type \mono{A} which are required and escaping the method \mono{m}.

\begin{figure}[h]
 \begin{scriptsize}
 \begin{lstlisting}
A m(int n, out A p2) {
	Contract.Memory.MemReq<A>(n + 5);
	Contract.Memory.Esc<A>(Contract.Memory.Return, 2);
	Contract.Memory.Esc<A>(Param, 1);
	Contract.Memory.BindEsc(Param, p2); 

	A a1 = new A();	
	Contract.Memory.DestEsc(Param);
	p2 = new A(); 	
	A a3 = m1(n);
	Contract.Memory.AddEsc(Contract.Memory.Return, Contract.Memory.Return);
	A a4 = m2(n);	
	return a4;
}
A m1(int m) {
	Contract.Memory.MemReq<A>(m + 1);
	Contract.Memory.Esc<A>(Contract.Memory.Return, 1);
	...
}
A m2(int k) {
	Contract.Memory.MemReq<A>(k);
	Contract.Memory.Esc<A>(Contract.Memory.Return, 2);
	...
}
 \end{lstlisting}
 \end{scriptsize}
\vspace{-.5em}
\caption{A simple example to illustrate the instrumentation process.}
\label{ex-baseinst}
\end{figure}

\begin{figure}[h]
 \begin{scriptsize}
 \begin{lstlisting}
A m(int m, out A p2) {
	Contract.Ensures(m_MemReq_A <= n + 5);
	Contract.Ensures(m_Esc_Return_A <= 2);
	Contract.Ensures(m_Esc_Param_A <= 1);

	m_MemReq_A = 0;
	m_Esc_Return_A = 0;
	m_Esc_Param_A = 0;

	m_MemReq_A += 1;
	A a1 = new A();
	
	m_MemReq_A += 1;	
 	m_Esc_Param_A += 1;	
	p2 = new A();
	
	int maxCalls_A = 0; 
	int sumCalls_A = 0;
	
	int call1_diff_A = (n + 1) - 1; // m1_MemReq_A - m1_Esc_A
	maxCalls_A = max(maxCalls_A, call1_diff_A);
	sumCalls_A += 1; // m1_Esc_A
	A a3 = m1(n);

	int call2_diff_A = n - 2;  // m2_MemReq_A - m2_Esc_A
	maxCalls_A = max(maxCalls_A, call2_diff_A);
	sumCalls_A += 2; // m2_Esc_A
	m_Esc_Return_A += 2; // m2_Esc_A	
	A a4 = m2(n);	

	m_MemReq_A += maxCalls_A + sumCalls_A;
	return a4;
}
 \end{lstlisting}
 \end{scriptsize}
\vspace{-.5em}
\caption{Instrumented version for our simple example.}
\label{ex-inst}
\end{figure}

The memory consumption annotations are transformed into corresponding ensure clauses stating that the associated counters are less than or equal to the specified bounds. For instance, the annotation \mono{Contract.Esc<C>(t, e)} is transformed into \mono{Contract.Ensures(\mono{m\_Esc\_t\_C} <= e)}. The same approach applies for \mono{MemReq} annotations.
For \mono{AddEsc(d, s)} annotations, the instrumentation consists in adding to the respective local counters the value of the callee counter.

One important aspect that we have overlooked so far is how to handle objects captured by a callee. 
As we mentioned previously, temporary objects created inside the method and not escaping its scope, can be recycled.
Again, consider the example in Figure~\ref{ex-baseinst}. There \mono{m}  invokes two methods: \mono{m1} and \mono{m2}, consuming $MR_1=m+1$, $E_1=1$ and  $MR_2=k$ and $E_2=2$ respectively. 
The objects escaping \mono{m1} and \mono{m2} must be handled by \mono{m} regardless of whether those objects end up escaping \mono{m}. Thus, they will be part of \mono{MemReq}. 
In addition, each method may create some temporary objects that can be recycled when they finish its execution. 
Thus, to compute \mono{MemReq} we can consider the  expression $\max\{MR_1+ E_2, MR_2+E_1\}$, representing this idea of reusing of objects\footnote{Our analysis is flow insensitive. This is because is not trivial to deal with the statement ordering in case of loops featuring conditions in the loop body.}. 
This expression can be rewritten as $\max\{MR_1-E_1, MR_2-E_2\}+E_1+E_2$ and binding the variables $m$ and $k$ with the call argument $n$ we obtain $\max\{n+1-1, n-2\}+1+2 = n+3$. We handle this computation by including two auxiliary counters per type called \mono{maxCalls} and \mono{sumCalls}. For instance, in the example of Figure~\ref{ex-inst} we use the variables \mono{maxCalls\_A} and \mono{sumCalls\_A} to accumulate the result of calls for the type \mono{A}.

For a method featuring  loops, the idea is to apply the same procedure  over every loop iteration to compute the maximum among all iterations. Again, we use one auxiliary counter for each expression that is computed during the loop. 
At the end of the loop these counters contain the expression representing the consumption for all the iterations. 
For instance, in Figure~\ref{ex1-inst} we show one fragment of the instrumented version of the method \mono{CreateFamily}. There, the variables \mono{maxCall\_Logger} and \mono{maxCall\_Person} compute the contribution of the calls in the loop for the objects of type \mono{Logger} and \mono{Person}. 

\begin{figure}[h]
 \begin{scriptsize}
 \begin{lstlisting}
Family CreateFamily(string lastName, string[] firstNames) {
	...
	int maxCall_Person = 0;	
	int sumCall_Person = 0;	
	int maxCall_Logger = 0;	
	int sumCall_Logger = 0;	

	for (int i = 0; i < firstNames.Length; ++i) {
		int call_diff_Person = 1 - 1; // AddMember_MemReq_Person - AddMember_Esc_This_Person
		maxCall_Person = max(maxCall_Person, call_diff_Person);
		sumCall_Person += 1; // AddMember_Esc_This_Person
		int call_diff_Logger = 1 - 0; // AddMember_MemReq_Logger - AddMember_Esc_This_Logger 
		maxCall_Logger = max(maxCall_Logger, call_diff_Logger);
		sumCall_Logger += 0; // AddMember_Esc_This_Logger

		family.AddMember(firstNames[i]);
	}

	CreateFamily_MemReq_Person += maxCall_Person + sumCall_Person;
	CreateFamily_MemReq_Logger += maxCall_Logger + sumCall_Logger;
	...
}
 \end{lstlisting}
 \end{scriptsize}
\vspace{-.5em}
\caption{Fragment of an instrumented version.}
\label{ex1-inst}
\end{figure}

\subsection{Verifying object lifetime annotations}
For the instrumentation and verification process we are assuming that object lifetime annotations, such as \mono{DestEsc}, are correct. To ensure they actually are, we include a lifetime annotations checker. 

To perform this verification we rely on a points-to and escape analysis capable of analyzing .NET programs~\cite{garber07iwaco}.  
A points-to and escape analysis is basically a tool that provides an abstraction of the program at a given program location.
A typical (finite) representation of a heap is given by a points-to graph (PTG). 
A PTG is a graph (L, N, E) where a node $n \in N$ represent a set of objects, in general all objects allocated in the same program point, 
 and a edge $(n1,f, n2)$  represents that one of the objects denoted by $n1$ may point-to one of the objects denoted by $n2$ using the field $f$.  $L$ is a mapping from local variables (including parameters) to nodes, which is used to express how the program accesses the heap using those variables. Using a PTG we can query if an object may be reachable from either some variable or from another object. This query can be easily solved by traversing the PTG starting from $L$, in case of a parameter, or a node $n$, in case of an object.  

\begin{figure}[ht]
\centering
\begin{minipage}{.4\textwidth}
\includegraphics[scale=.6]{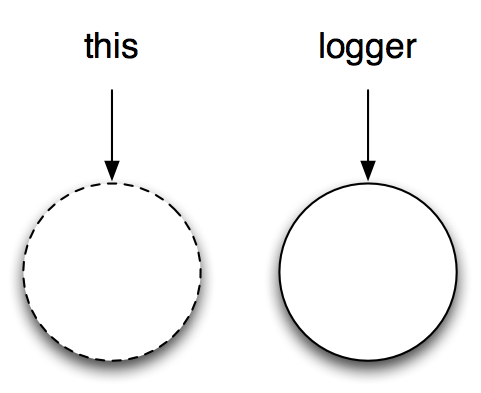}
\end{minipage}
\qquad 
\begin{minipage}{.4\textwidth}
 \includegraphics[scale=.6]{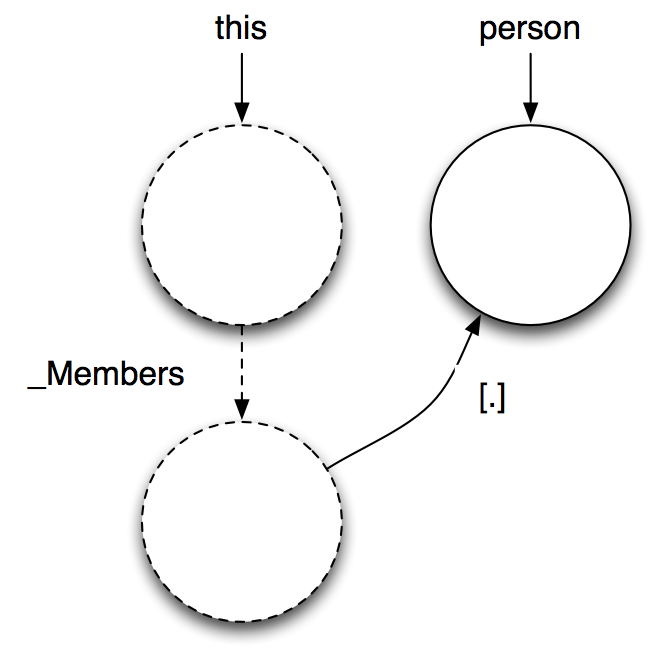}
\end{minipage}
\caption{PTGs for the constructor of \mono{Person} and the \mono{AddMember} method of \mono{Family}.}
\label{fig:ptgs}
\end{figure}

For instance, consider the Figure~\ref{fig:ptgs}. The first PTG corresponds to the exit point of the \mono{Person} constructor. A dotted circle represents a parameter or load node meaning objects created outside the scope of the method under analysis. In this case, the node models the object represented by the parameter \mono{this}. A solid circle represents an object created during the execution of the method (or some of its callees). In this case, it models the \mono{logger} object. The second PTG corresponds to the exit point of \mono{AddMember}. In this case we can see that this method does not allocate objects but makes the object pointed by \mono{person} reachable from \mono{this}. 

Our analysis computes a PTG for each annotated method. This PTG is an abstraction of the program heap visible from each method at the end of its execution. 
To verify that objects not annotated are in fact captured by the method, we check whether the node in the PTG, representing objects created at that program point, is not reachable from the global scope, method parameters or the returned object.  

For \mono{DestEsc} we check if the object escapes and it is reachable through the path expression associated with the tag (i.e.,  by using \mono{BindEsc}, as we do in the example of Figure~\ref{ex-baseinst}). The correctness of the lifetime information associated to the annotation \mono{AddEsc}, where the objects are assigned to the corresponding tag, is also verified in a similar fashion.

The points-to analysis we use builds a conservative abstraction. That means that it may produce a false alarm and state that an object may escape when it does not.
To overcome this issue, we provide the annotation \mono{DestLocal} for new statements which tells the verification engine to ignore the result of the escape analysis for that allocation.


\section{Discussion} \label{sec:barvinok}

\subsection{Verifying complex contracts}

\textsc{Clousot} does a very good job in performing automatic verification of contracts, being able to check method featuring loops without demanding loop invariants. 
However, it has some limitations when dealing with the complex arithmetic required for a quantitative analysis. According to our experiments the current version of \textsc{Clousot} is restrained to contracts having linear integer expressions.
 
One possible approach to improve our analysis is relying on theorem provers. Fully automatic and complete decision procedures for non-linear constraints is impossible, one needs to sacrifice completeness or automation. Nevertheless, modern SMT solvers like \textsc{Z3}~\cite{z3} started to provide some automatic (but incomplete) support for polynomials. Thus, it can be possible to use \textsc{Z3} to verify this resource consumption expression. The problem with this approach is that the user would need to provide loop invariants describing how the consumption is evolving during loops. This task could be very complex and error-prone.

Another approach is  trying directly to infer the consumption in those cases.   In~\cite{garber08ismm,garber11cpe} we propose a technique to automatically infer memory consumption bounds on Java like programs.
Essentially the technique works as follows: given a method featuring a loop (with possible several nested loops) including a \mono{new} statement  
and a predicate describing its iteration space (i.e. a linear restriction describing the relation between the loop inductive variables and parameters),  we can obtain a parametric upper-bound of the number of times the  \mono{new} statement is executed. This upper bound is obtained by counting the number of solutions of that iteration space. 
In a similar fashion, we can deal with polynomial temporary and escaping consumptions by applying respectively a symbolic maximization and sum operations over the iteration space~\cite{garber12scp}. 
\textsc{Barvinok}~\cite{clauss2009symbolic} is a tool\footnote{\small Available at: \url{http://freshmeat.net/projects/barvinok}.} capable of manipulating parametric integer sets and relations.  It provides functionality to \emph{count} the number of elements of these sets and for performing \emph{maximization} and \emph{sum} on polynomials over these sets. 

For those methods whose consumption is beyond the capabilities of \textsc{Clousot} we may use this approach. 
The price to pay to obtain more precision is the need of a new annotation to specify iteration spaces inside loops: \mono{IterationSpace}. 
Although this increases the annotation burden,  the gain is considerable since it makes possible the verification (and inference) of polynomial consumption. 
Notice that iteration spaces are a set of linear constraints, amenable to  be checked with \textsc{Code Contracts} as well and could even be inferred using \textsc{Clousot} ability for inferring loop invariants.

Figure~\ref{ex3} shows a method with a loop and a nested loop inside it. In this case \textsc{Clousot} would not be able to verify the contract. 
However, following the technique presented in~\cite{garber08ismm}, using  \textsc{Barvinok} with the aid of~\mono{IterationSpace} annotations in the loops, we can determine the exact number of times that the method call on line~\ref{ex3:new-2loop} is executed and instruct the engine with new knowledge.

\begin{figure}[h]
\begin{scriptsize}
\begin{lstlisting}
public Family CreateBigFamily(int n) {
    Contract.Requires(n > 0);
    ...
    Contract.Memory.Esc<Person>(Contract.Memory.Return, n * (n + 1) / 2);
    ...
    Contract.Memory.DestEsc(Contract.Memory.Return);
    Contract.Memory.AddEsc(Contract.Memory.Return, Contract.Memory.This);
    Family family = new Family("Doe", n * (n + 1) / 2);

    for (int i = 1; i <= n; i++) {
        Contract.Memory.IterationSpace(1 <= i && i <= n);

        for (int j = 1; j <= i; j++) {
            Contract.Memory.IterationSpace(1 <= j && j <= i);

            Contract.Memory.AddEsc(Contract.Memory.Return, Contract.Memory.This);
            family.AddMember("John"); (*@ \label{ex3:new-2loop} @*)
        }
    }

    return family;
}
\end{lstlisting}
\end{scriptsize}
\vspace{-.5em}
\caption{Using \mono{IterationSpace} to assist the prover.}
\label{ex3}
\end{figure}

\subsection{From objects to real memory consumption}

Our analysis is focused in quantifying live objects rather than computing the exact memory consumption of a program. 
Computing real consumption requires to solve challenges posed by particularities of .NET or Java virtual machines like their garbage collectors (GC), representation of arrays and memory fragmentation. 
Nevertheless, we believe we are still targeting the main problem which is related to the number of objects allocated by a program and how many of them may be released during the execution, in order to reuse the memory. 

Our annotations allow the programmer to quantify memory consumption by tags. The tags enable the verifier to take into account that there is a memory manager that collects objects when they are no longer needed. 
As mentioned, we decided to allow memory reclaiming at the method scope. This is, of course, an approximation of how real garbage collection may perform\footnote{Actually, the analysis over-approximate the behavior of an \emph{ideal} garbage collector. That is a GC that collects objects as soon as they become unreachable.} and other fine-grained mechanisms need to be investigated. 
We also need to investigate how to specify the memory consumed by the VM itself, which is not visible by looking at the application code.
 
\subsection{About annotations and information hiding}

The annotation language enables to aggregate information about objects according to their types. This can be particularly useful when we need approximate real memory consumption as we can use the size of each type to know its required space in memory (modulo some possible fragmentation).

However, the information about types jeopardizes information hiding. Suppose we have a method $m$ that requires some auxiliary objects of type~\texttt{T} perform a task. These objects do not escape the method, but $m$ needs to declare that it requires $n$ objects of type~\texttt{T} to run. This information will be propagated to the caller and even to their parents (even the main method!), making the type~\texttt{T} visible to other fragments of code, and thus breaking information hiding. 
In our running example, this phenomena occurs with the auxiliary object of type \mono{Logger} used in the \mono{Person} constructor, whose type is propagated to its callers.

One possible solution to overcome this problem is to just quantify objects, regardless of their type, as can be seen in Figure~\ref{exMCallObjs}. This would make the analysis less useful for computing real memory bounds, but still can be used to compare different  implementations of algorithms in terms of their memory usage. Another choice is qualifying the consumption using other kinds of units of measuring, like number of fields, bytes, etc. We need to investigate further to understand which is the right balance between granularity, verifiability, understandability, utility and modularity of contracts.

\begin{figure}[ht]
\begin{scriptsize}
\begin{lstlisting}[escapeinside={@}{@}]
Family CreateFamily(string lastName, string[] firstNames) {
	Contract.Memory.MemReq<object>(2 + 2 * firstNames.Length );
	Contract.Memory.Esc<object>(Contract.Memory.Return, 1 + 2 * firstNames.Length);
	...
}
\end{lstlisting}
\end{scriptsize}
\vspace{-.5em}
\caption{Contract hiding type information.}
\label{exMCallObjs}
\end{figure}

\section{Implementation details} \label{sec:implementation}
We implemented our prototype tool as a \textsc{Visual Studio} extension\footnote{\small Available at: \url{http://lafhis.dc.uba.ar/resourcecontracts}.} that lets developers write memory consumption contracts as  they do with \textsc{Code Contracts} and verify them using its static verifier or run-time checker.

The only prerequisite for the plug-in is having \textsc{Code Contracts} installed, all the other tools used by the memory contracts checker are packaged in the plugin and relies on the  Common Compiler Infrastructure (\textsc{CCI})~\cite{CCI} for code analysis and instrumentation.
 
 \begin{figure}[ht]
 \begin{center}
 \includegraphics[scale=0.5]{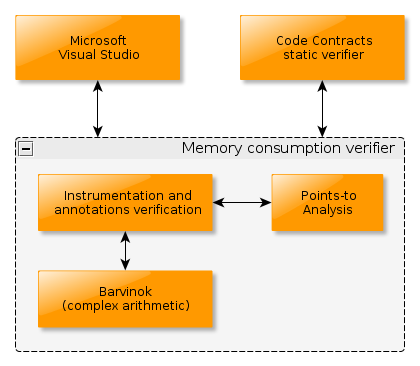}
 \end{center}
 \caption{General architecture of the tool.}
 \label{fig:arq}
 \end{figure}
 
\begin{figure}[hb]
 \begin{center}
 \includegraphics[scale=0.6]{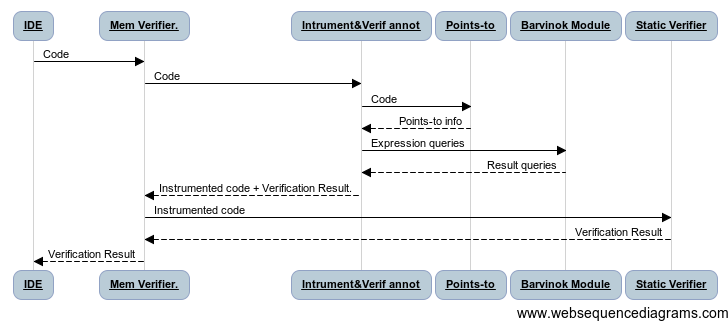}
 \end{center}
 \caption{Interaction between main modules.}
 \label{fig:seq-diag}
 \end{figure}

Figure~\ref{fig:arq} shows the general architecture of the tool and Figure~\ref{fig:seq-diag} presents a sequence diagram showing the interaction made by the main module in order to verify resource contracts.
The component labeled \textit{Memory consumption verifier} has the same interface as the \textsc{Code Contracts} verifier, so it can replace it when \textsc{Visual Studio} invokes it.  Internally, the \textit{Memory consumption verifier} uses the described algorithms and tools to do the instrumentation and verification, then it invokes the \textsc{Code Contracts} verifier and returns to \textsc{Visual Studio} the verification results.

The \textsc{Code Contracts} static checker is invoked by \textsc{Visual Studio} after each compilation. 
In order to transform the code before checking it, we use a wrapper that performs the required instrumentation and then invokes the actual checker. 

\section{Related work} \label{sec:relatedwork}
Recently, there were relevant advances in resource analysis for imperative and functional programs~\cite{chin05SAS,BarthePS05,garber08ismm, garber11cpe, albert12scp,he2009memory,gulwani2010reachability,hoffmann2010amortized}. Most of them are aimed at inferring of resource consumption bounds rather than verifying them.
Here will  briefly refer to some of them which are focused in verification of annotated programs. 

The work in~\cite{chin05SAS}  proposes a type system to statically check linear size annotations (Presburger's formulas) in a functional fragment of a Java-like language. This approach allows  specifications of the number of preexistent objects released  by a method, but it requires  complex annotations in order to specify potential aliasing relations between expressions in the language. 
We prefer a coarse grained approach demanding less and easier to infer annotations.

Closer to our approach,~\cite{BarthePS05} defines an annotation language based on JML that can be used to annotate Java bytecode. This language is limited and does not contemplate the specification of lifetime information. In~\cite{he2009memory} the authors present a verification system for C-like programs using recursion as the only iteration mechanism. Similar to ours, they use contracts and program instrumentation techniques using a non-specialized verifier. Their system supports \mono{free} statements which in principle enables a more precise reasoning. However, according to our experience, verifying non-linear consumption in those systems is extremely hard because of the need of machinery capable of dealing with lower and upper bounds. 

Atkey~\cite{Atkey2011} presents a technique to verify imperative pointer-manipulating languages by embedding a resource logic, which is   an extension of separation logic. 
The technique is inspired in the concept of amortized resource analysis~\cite{hofm03}. 
Essentially, the logic enables the specification of  input/output potentials for each method. Input potentials are used to represent the required resources to run the method and output potentials represent the resources that are warranted to be available at the end of its execution. 
He also describes a proof search procedure that allows generated verification conditions to be discharged while using linear programming to infer consumable resource annotations. 
In contrast with our approach, his language has an explicit construct to manually deallocate objects and their current logic only allows linear restraints.  

More recently, Albert and collaborators~\cite{AlbertBGHR12} presented an approach to verify memory consumption contracts using a verifier called \textsc{Key}~\cite{ahrendt2005key}. 
Essentially, they propose to use their tool~\textsc{Costa}~\cite{albert12scp}  to infer annotations about memory consumption and then apply the solver \textsc{Key} to verified them. 
In this case they need to rely on the power of the verifier to deal with the non-linearity of the inferred bounds. 
\textsc{Costa} may help by providing some annotation but still the user may need to provide invariants and annotations about the shape and lifetime of objects. 
In contrast, in our approach we are more focused in automating the process by providing an automatic escape analyzer, using an static analyzer which does not require invariants, and using an external tool to deal with non-linear expressions.

\section{Conclusions and future work} \label{sec:conclusions}
In this work we presented an extension of  \textsc{Code Contracts} language to specify and verify the memory consumption of .NET programs.
The tool integrates with \textsc{Visual Studio} enabling autocompletion, inline  documentation, static verification and run-time checking as \textsc{Code Contracts} does.                                                  

As a future work, we would like to address more complex and larger programs. 
To do so we plan  to enhance the usability of the tool by automatically inferring quantitative and lifetime annotations.
In this setting developers would only need to specify complex or hard-to-infer annotations, not worrying about annotations that can be easily inferred. 
In this matter, we plan to port our previous work on inference of memory consumption for Java~\cite{garber08ismm} to .NET and extend other tools capable of inferring resource usage (e.g.,~\cite{gulwani2010reachability}) in order to make them capable of dealing with dynamic memory usage.

We plan to formally prove that the instrumentation technique preserves the behavior of the original program in terms of memory consumption.  

Our current annotations depends on an external points-to analysis that is used to approximate escape analysis information. This analysis must be executed before our verification process. 
We would like to extend our annotation language to allow also to express directly lifetime properties (while maintaining information hiding)  and being able to modularly verify programs without requiring such external tool.

\section*{Acknowledgments}
This work has been partially funded by CONICET, UBACyT-20020110200075/20020100100813, MinCyT PICT-2010-235/2011-1774/2012-0724, CONICET-PIP 11220110100596CO, MINCYT-BMWF AU/10/19, INRIA Associated Team ANCOME, and LIA INFINIS and MEALS 295261.
\bibliographystyle{eptcs} 
\bibliography{garbervetsky}

\end{document}